\documentclass[12pt]{iopart}

\usepackage{graphicx}
\usepackage{amssymb}
\usepackage{color}
\usepackage{verbatim}
\usepackage{array}
\usepackage{graphicx}
\usepackage{epsfig}
\usepackage{bm}
\usepackage[ragged]{sidecap}
\usepackage{dsfont}
\usepackage{ulem}

\newcommand{\ds}{\displaystyle}
\newcommand{\ddsum}[1]{{\displaystyle \sum_{ #1 }}}

\newcommand{\ddsumd}[2]{{\displaystyle \sum_{ #1 }^{ #2 }}}

\newcommand{\supercomas}[1]{``#1''}

\def\bra#1{\mathinner{\langle{#1}|}}
\def\ket#1{\mathinner{|{#1}\rangle}}
\newcommand{\braket}[2]{\langle #1|#2\rangle}

\begin{document}

\title{Underlining some limitations of the statistical formalism in quantum mechanics}

\author{F Fratini$^{1}$ and A G Hayrapetyan$^{2}$}
\address{$^1$
Ruprecht-Karls-Universit\"at, 69120 Heidelberg, Germany}
\ead{fratini@physi.uni-heidelberg.de}
\vspace{0.2cm}
\address{$^2$
Max-Planck-Institut f\"ur Kernphysik, Postfach 103980, 69029 Heidelberg, Germany}

\begin{abstract}
We show that two chosen ensembles of spin states, which are 
differently prepared but are described by the same density matrix
in quantum mechanics, do not fully share the same measurable characteristics.
One characteristic on which they differ is shown to be the variance of the spin along a given direction.
We conclude that the statistical description of an ensemble of states as given
by its density matrix, although sufficient in many cases, should be considered incomplete, as it does not fully describe
the measurable characteristics of the ensemble. 
A discussion a posteriori on the problem is provided.
\end{abstract}



\section{Introduction}
Density matrices are naturally used in quantum mechanics
for describing physical systems whose quantum states are not equal and to which a single wavefunction
cannot be therefore assigned. These systems are said to
be in a mixed state and are described, within the density matrix formalism, by means of an incoherent superposition of pure states.
By definition, incoherent superposition means that the probability of getting a certain experimental result, 
as a consequence of a measurement on the ensemble, is given by taking the 
average of the results which are obtained for the pure states
the ensemble is made of, where the weights are the population fractions (or numbers) of the pure states.
Upon this principle, the density matrix formalism has been introduced into quantum mechanics 
in the late 20's by Von Neumann, Dirac and Landau, who developed and applied it
to quantum information theory, statistical thermodynamics and wave mechanics \cite{Neu1927, Dir1929, Lan1927}.\\
Density matrices have been extensively applied for studying numerous
Physics fields. Few examples may include quantum phase transitions \cite{WU}, many-particle systems \cite{PER}, entanglement 
measures \cite{WOOTERS, FRA} and scattering processes \cite{ND}. 
Thorough reviews of the density matrix formalism and its applications can be found
in Refs.~\cite{FANO, McW1960, BLUM, BAL}.\\
The evaluation of density matrices for complex systems may present 
mathematical difficulties. 
For a harmonic oscillator in a thermostat, for example,
the density matrix can be built only by
performing a double, both quantum and statistical, average over the coherent states.
Such a density matrix was successfully derived in 1932 by Bloch \cite{Bloch:32}, who used a rather cumbersome mathematical
apparatus, and more recently 
by Avakyan {\it et al.} \cite{Avakyan:07}, by using Glauber's coherent
states \cite{Glauber:63, Glauber:63b}, which allow a more concise derivation.\\
In the present contribution,
we show that ensembles of states, which are represented by the same density operator in quantum mechanics,
i.e. by the same density matrix, may not fully share the same characteristics to measure. 
Consequently, we remark that the description of an ensemble given
by its density operator, or by its density matrix, 
although sufficient in many cases, should be considered incomplete, as it does not fully describe
the measurable characteristics of the ensemble. \\
The paper is structured as follows. 
In section \ref{sec:ens_mea} we define two ensembles of spin states which will be used 
throughout the paper. In 
section \ref{sec:one} we derive, by using primary principles of quantum mechanics and statistics, 
the expectation value and the variance of the spin along a chosen direction for both ensembles.
In section \ref{sec:two}, we briefly recall the rudiments of the theory of statistical quantum mechanics and we 
apply such theory to the previously defined ensembles. The expectation value and the variance of the spin along the same chosen direction
are then calculated once again, for both ensembles, by using statistical quantum mechanics. The variances of the spin 
are found to differ from the ones presented before.
A discussion a posteriori on the problem is presented in section \ref{sec:disc}, where we show that 
the obtained discrepancy has to be attributed to
the inapplicability of the statistical formalism of quantum mechanics to the evaluation of the variance.
A brief summary is given in section \ref{sec:sum_conc}.

\section{Ensembles definitions and observables to investigate}
\label{sec:ens_mea}
The ensembles we consider are made of non-interacting spin-1/2 particles. To some 
extent, spin-1/2 particles are the most representative particles in quantum mechanics, since any measurement of their spin along 
whichever axis can result in only two possible outcomes: either $+\hbar$/2 or $-\hbar$/2. \\
After having defined any system 
of coordinate axes, we consider two ensembles of particles defined as follows:
\begin{itemize}
\item ensemble $\mathcal{A}$: $N$ particles whose spin states are defined along the $\hat x$ axis, 
$N/2$ of which are eigenstates of the spin operator $\hat{S}_x$ with eigenvalue $+\hbar$/2, while the rest $N/2$ are eigenstates of 
the same operator with eigenvalue $-\hbar$/2 ;\\
\item ensemble $\mathcal{B}$: $N$ particles whose spin 
states are defined along the $\hat z$ axis, $N/2$ of which are eigenstates of the spin operator $\hat{S}_z$ with eigenvalue $+\hbar/2$,
while the rest 
$N/2$ are eigenstates of the same operator with eigenvalue $-\hbar/2$.
\end{itemize}
Both $\mathcal{A}$ and $\mathcal{B}$ are totally unpolarized ensembles.
In the language of statistical quantum mechanics, which will be encountered
in section \ref{sec:two}, they are also said to be in a maximally mixed state.\\
We shall focus
on deriving, for both ensembles, the expectation value and the variance of the spin along the $\hat x$ direction. 
Both of them are measurable quantities. \\
We recall that 
the expectation value of some discrete variable $y$ distributed according to a certain probability 
function $f(y)$ is defined as \cite{StatData}
\begin{equation}
\textrm{E}(y)=\ddsum{i}f(y_i)\, y_i \, ,
\label{exp_val}
\end{equation}
where $i$ labels the values that the variable $y$ is allowed to assume.\\
The variance of $y$, which is a measure of how widely $y$ is spread about its mean value, is obtained as \cite{StatData}
\begin{equation}
\begin{array}{l c l}
\textrm{Var}(y)&=& \textrm{E}\left(    \left(y - \textrm{E}(y)\right)^2   \right)\\[0.3cm]
&=& \textrm{E}(y^2) - \left(\textrm{E}(y)\right)^2   \, .
\end{array}
\label{vardef}
\end{equation}
Normally, the distribution function $f(y)$ is experimentally measured by performing
many measurement of the observable $y$.	However, in many cases $f(y)$ 
can be theoretically derived, so that predictions
for the expectation value and the variance can be provided. \\
If we have
some operator $\hat{O}$ and its discrete spectrum of eigenstates $\ket{o_i}$ which satisfy the equation
\begin{equation}
\ds \hat{O}\ket{o_i} = o_i \ket{o_i} \, ,
\label{eigen}
\end{equation}
in quantum mechanics it is postulated that
the expectation value of the variable $o$ over a certain state $\ket{\beta}$ can be theoretically obtained by
calculating \cite{DIR, SAK}
\begin{equation}
\ds \textrm{E}(o)= \bra{\beta} \hat{O} \ket{\beta} = \ddsum{i} \Big| \braket{o_i}{\beta} \Big|^2 o_i\;,
\label{expvaluewave}
\end{equation}
where the identity 
\begin{equation}
\ddsum{i}  \; \ket{o_i}\bra{o_i} = \hat{\mathds{1}} 
\end{equation}
and the normalization equation
\begin{equation}
\braket{o_i}{o_j} = \delta_{i,j}
\end{equation}
have been used in the last step of equation (\ref{expvaluewave}). 
By comparing equation (\ref{expvaluewave}) with equation (\ref{exp_val}), we notice that the probability function is evidently given by
\begin{equation}
\ds f(o_i)= \Big| \braket{o_i}{\beta} \Big|^2\, .
\label{distrfun}
\end{equation}
In the next section, we shall show that, while the expectation values of the spin along the $\hat{x}$ direction coincide for the 
ensembles $\mathcal{A}$ and $\mathcal{B}$, the variances do not.

\section{Expectation value and Variance for the ensembles}
\label{sec:one}
In this section, we derive the expectation value and 
the variance of the spin along the $\hat x$ direction for the ensembles of states $\mathcal{A}$ and $\mathcal{B}$
defined in 
section \ref{sec:ens_mea}. 
We will work separately on the two ensembles, without turning to the quantum statistical description.	
For this purpose, we can pragmatically think of using, for example, 
a Stern and Gerlach (SG) apparatus with the magnetic field along the $\hat x$ direction \cite{SG}. The SG apparatus would measure 
the spin of each single particle of the ensemble, so that the total spin of the ensemble would then be obtained as the
sum of all the single particle spin measurements. \\
Since the present manuscript is theoretically oriented, we will consider
the SG apparatus as ideal, without any experimental limitation.
We will furthermore denote with $\ket{S_x, \pm 1}$ and $\ket{S_z, \pm 1}$ 
the eigenstates of the operators 
$\hat{S}_x$ and $\hat{S}_z$ respectively. The corresponding eigenvalues are
$\pm \hbar /2$ in both cases.

\subsection{Ensemble $\mathcal{A}$}
Due to the characteristics chosen for the ensemble $\mathcal{A}$, the SG apparatus
would exactly
separate the particles flux into two equal parts, or, which is the same, it will
measure $N/2$ particles 
having spin along $+\hat x$ direction and $N/2$ particles having spin along $-\hat x$ direction. 
The probability of registering the outcome $\pm \hbar/2$, when the spin along $\hat x$ is measured on the state $\ket{S_x, \pm 1}$, is in fact
certainly 1.\\
Consequently, the total spin of the ensemble $\mathcal{A}$, as measured along the $\hat x$ direction, will be always $0$, 
in any measurement which is carried out on the ensemble, independently of the number of particles contained in $\mathcal{A}$. 
Since, as already recalled in section \ref{sec:ens_mea}, the variance is a measure of
how widely the single measurements are spread about the mean value,
the variance for the spin measurement on the ensemble $\mathcal{A}$ is also vanishing.\\
We can summarize what previously stated by writing
\begin{equation}
\left.
\begin{array}{l c l}
\textrm{E}(S_x)_{\mathcal{A}}& = & 0 \\
\textrm{Var}(S_x)_{\mathcal{A}} & = & 0
\end{array}
\right\}\Rightarrow S_x^{\mathcal{A}} = 0 \pm 0 ~,
\label{predONEA}
\end{equation}
where the subscript $\mathcal{A}$ denotes the ensemble on which the expectation value and the variance are evaluated.
$S_x^{\mathcal{A}}$ represents our theoretical prediction for the measurement 
of the spin along the $\hat x$ direction on the ensemble $\mathcal{A}$.

\subsection{Ensemble $\mathcal{B}$}
Differently from the ensemble $\mathcal{A}$, each particle of the ensamble $\mathcal{B}$,
independently the one from the other, has probability one-half of being registered with spin along $+\hat x$ direction and one-half of 
being registered with spin along $-\hat x$ direction. 
This obvious statement can be easily proved by writing the spin states which are defined along the $\hat z$ direction as \cite{SAK}
\begin{equation}
\ket{S_z, \pm 1} = \frac{1}{\sqrt{2}}\left( \ket{S_x, + 1} \pm \ket{S_x, - 1} \right) ~,
\label{five}
\end{equation} 
and by noticing that the probability of registering the outcome $\pm\hbar/2$, in a measurement of the spin along the
$\hat x$ axis of the state in equation (\ref{five}), can be obtained by using equation (\ref{distrfun}):
\begin{equation}
\left|\braket{S_x, \pm 1}{S_z, \pm 1}\right|^2 = \frac{1}{2} ~.
\end{equation} 
In the light of what said above, the measurement outcomes from the SG apparatus must follow the binomial distribution. This 
means that $N/2$ particles are expected to be measured with 
spin along $+\hat x$ direction and $N/2$ particles with spin along $-\hat x$ direction. 
Being a binomial distribution, the variance of these expected particles numbers
can be easily calculated to be $N/4$ \cite{StatData}. 
If we adopt the standard deviation ($\sigma=\sqrt{\textrm{Var}}$) as indeterminacy of the measurement, we may write that the 
number of particles measured with spin equal to $+\hbar/2$ (as well as the
number of particles measured with spin equal to $-\hbar/2$) must be
\begin{equation}
\frac{N}{2} \pm \frac{\sqrt{N}}{2} ~.
\label{npart}
\end{equation}
Consequently, the total spin of the ensemble $\mathcal{B}$ along the $\hat x$ direction, as measured by the SG apparatus,
must be
\begin{equation}
\hat{S}_x^{\mathcal{B}} = \frac{\hbar}{2}\left( 0 \pm \frac{\sqrt{N}}{2}\; 2\right) =
0 \pm \frac{\hbar \sqrt{N}}{2}  ~.
\label{Ex}
\end{equation}
The factor \supercomas{2} which multiplies the standard deviation in equation (\ref{Ex}) arises from the fact that `\,not detecting
a particle with spin $+\hbar/2$' results in `\,detecting that particle with spin $-\hbar/2$'.\\
Summarizing what obtained for the ensemble $\mathcal{B}$:
\begin{equation}
\left.
\begin{array}{l c l}
\textrm{E}(S_x)_{\mathcal{B}}& = & 0 \\
\ds \textrm{Var}(S_x)_{\mathcal{B}} & = & \frac{\hbar^2 N}{4}
\end{array}
\ds \right\}\Rightarrow S_x^{\mathcal{B}} = 0 \pm \frac{\hbar\sqrt{N}}{2} ~,
\label{predONEB}
\end{equation}
where $S_x^{\mathcal{B}}$ represents our theoretical prediction for the measurement 
of the spin along the $\hat x$ direction on the ensemble $\mathcal{B}$.

\medskip

We can certainly conclude that the two ensembles $\mathcal{A}$ and $\mathcal{B}$ are not 
equal, as they show a measurable difference, which is the variance of the spin along the $\hat{x}$ direction.
The latter can be naturally measured in experiments
by extracting the width of the measured spin distribution.\\
Since only few solid principles of quantum mechanics and of statistics have been used up to now,
the above predictions are considered reliable: any measurement is reasonably expected to agree with them.

\section{Quantum statistical description}
\label{sec:two}
The overall state of an admixture of particles with different states is formally described in statistical quantum mechanics by means of 
the so-called \supercomas{density operator}. 
The {\it normalized} density operator for an ensemble of states reads \cite{BLUM, BAL, SAK}
\begin{equation}
\hat{P}=\ddsum{i} W_i \ket{a_i}\bra{a_i} \, ,
\label{DensOp}
\end{equation}
where $W_i$ is the statistical weight of the state $\ket{a_i}$, or, more practically, is the fraction of particles of the 
ensemble that share the same pure state $\ket{a_i}$. 
If $W_i = \delta_{ij}$, then the ensemble is said to be in the pure state $\ket{a_j}$. In any other case, 
the ensemble is said to be in a mixed state. If $W_i=W_0$, where $W_0$ is constant for all $i$, the ensemble is said to
be in a maximally mixed state.\\
The density matrix which describes the ensemble, in a given representation $\ket{b_i}$, is obtained as
\begin{equation}
\rho_{ij}=\bra{b_i}\hat{P}\ket{b_j} \, .
\label{DensMatr}
\end{equation}
Since it does not make any substantial difference between 
describing an ensemble of states by means of its density matrix or by means of its density operator, 
we choose to deal, in the following, with density operators. \\
The density 
operator for the ensemble $\mathcal{A}$ can be
explicitly written by applying the definition given in section \ref{sec:ens_mea}
to equation (\ref{DensOp}):
\begin{equation}
\begin{array}{l c l}
\hat{P}_{\mathcal{A}}&=&0.5 \ket{S_x,+1} \bra{S_x, +1} + 0.5 \ket{S_x, -1} \bra{S_x, -1}   \, .
\end{array}
\label{eq1}
\end{equation}
Similarly, the density operator for the ensemble 
$\mathcal{B}$ reads
\begin{equation}
\begin{array}{l c l}
\hat{P}_{\mathcal{B}}& = &0.5 \ket{S_z,+1} \bra{S_z, +1} + 0.5 \ket{S_z, -1} \bra{S_z, -1}\\ 
 & = & 0.5 \ket{S_x,+1} \bra{S_x, +1} + 0.5 \ket{S_x, -1} \bra{S_x, -1} ~,
\end{array}
\label{eq2}
\end{equation}
where, in the last step, we made use of equation (\ref{five}),
which link spin states defined along the $\hat x$ axis with spin states defined along the $\hat z$ axis \cite{SAK}.\\
By comparing equation (\ref{eq1}) with equation (\ref{eq2}), we 
notice that the density operators related to the ensembles $\mathcal{A}$ and $\mathcal{B}$ are analytically the same. 
Since 
the density operator is supposed to contain the whole information of the ensemble \cite{BLUM, BAL}, we should 
conclude that $\mathcal{A}$ 
and $\mathcal{B}$ show exactly the same polarization features, i.e. they are identical from the point of view of any 
polarization measurement. However, 
such a conclusion leads to a contradiction. As remarked in section \ref{sec:one}, the ensembles $\mathcal{A}$ and $\mathcal{B}$ show at least
one measurable difference, which is the Variance of the spin along the $\hat x$ direction.
It can be easily understood that the reason of this difference lies 
on the different physical characteristics which have been chosen a priori to characterize the two ensembles,
i.e. on the information concerning the preparation of the ensembles. \\
To complete the analysis, 
we derive, similarly to 
equations (\ref{predONEA}) and (\ref{predONEB}), the expectation value and the variance that the quantum statistical formalism provides for the ensembles.
For this purpose, we need first to recall that, in such a formalism, the expectation value of some variable
$o$ is given by \cite{BLUM, BAL, SAK}
\begin{equation}
\textrm{E}(o)= \textrm{Tr}[\hat{P}\hat{O}] ~,
\label{eq3}
\end{equation}
where $\hat{O}$ is the quantum mechanical operator related to $o$ as in equation (\ref{eigen}), 
$\textrm{Tr}$ denotes the Trace over any set of quantum states and $\hat{P}$ is the 
density operator which describes the ensemble on which the variable $o$ is measured. \\
By combining equations (\ref{vardef}), (\ref{eq1}), (\ref{eq2}) and (\ref{eq3}), we obtain:
\begin{equation}
\begin{array}{l}
\left.
\begin{array}{l c l c l}
\textrm{E}(S_x)_{\mathcal{A}}&=& \textrm{E}(S_x)_{\mathcal{B}}& = & \textrm{Tr}[\hat{P}_{\mathcal{A}}\hat{S}_x] 
= 0\\[0.2cm]
\textrm{Var}(S_x)_{\mathcal{A}}& =&\textrm{Var}(S_x)_{\mathcal{B}} 
&=& \textrm{Tr}[\hat{P}_{\mathcal{A}}\hat{S}_x^2]- \left(\textrm{Tr}[\hat{P}_{\mathcal{A}}\hat{S}_x]\right)^2 \\
&=&\ds\frac{\hbar^2}{4}
\end{array}
\right\}\\[0.2cm]
\ds \qquad \Rightarrow S_x^{\mathcal{A}} = S_x^{\mathcal{B}} = 0 \pm  \frac{\hbar}{2}~.
\end{array}
\label{predTWOAB}
\end{equation}
These results are in decisive contrast with equations (\ref{predONEA}) and (\ref{predONEB}). \\
We may now notice that the definition of the normalized density operator in equation (\ref{DensOp})
does not account for the number of 
particles that the ensemble contains. By virtue of this, any measurable quantity that depends on such number, like the variance
of the spin measurement over the ensemble along whichever axis, cannot be adequately described 
by using the normalized density operator. Consequently, we are lead to think that
the usage of the {\it not normalized} density operator for the two ensembles, which reads \cite{McW1960}
\begin{equation}
\hat{P}_{\mathcal{A}}=\hat{P}_{\mathcal{B}}=\frac{N}{2}\Big( \ket{S_z, +1}\bra{S_z, +1} + \ket{S_z, -1}\bra{S_z, -1}\Big) \, ,
\label{P_AB}
\end{equation}
might benefit the calculation. 
Indeed, by using equations (\ref{eq3}) and (\ref{P_AB}), the results we get in this case are
\begin{equation}
\begin{array}{l}
\left.
\begin{array}{l c l c l}
\textrm{E}(S_x)_{\mathcal{A}}&=& \textrm{E}(S_x)_{\mathcal{B}}& = & 0\\[0.2cm]
\textrm{Var}(S_x)_{\mathcal{A}}& =&\textrm{Var}(S_x)_{\mathcal{B}} & = & 
\ds\frac{\hbar^2 N}{4}
\end{array}
\right\}\\[0.2cm]
\ds \qquad \Rightarrow S_x^{\mathcal{A}} = S_x^{\mathcal{B}} = 0 \pm  \frac{\hbar\sqrt{N}}{2} ~,
\end{array}	
\label{predTWOABTWO}
\end{equation}
which are correct for the ensemble $\mathcal{B}$, as they match equation (\ref{predONEB}), but still not correct for 
the ensemble $\mathcal{A}$, as they do not match 
equation (\ref{predONEA}).\\
Since the density operators for the ensembles $\mathcal{A}$ and $\mathcal{B}$ are the same while the 
correct results for the variance of the spin along the $\hat{x}$ direction
are not, it is evidently not possible to recover the right description for both ensembles within the
quantum statistical formalism.

\section{Discussion}
\label{sec:disc}

To analyze the discrepancy which has been found in the previous sections, we first start by recalling how the density operator 
is normally introduced into quantum mechanics.\\
The leading idea which brings to the formulation of the density operator is the following.
Given an ensemble $\chi$ made of
$N$ sub-ensembles, where each one of the latter
is 
composed by particles
which share the same quantum state $\ket{\beta_i}$ ($i=1,...,N$),
the expectation value of a certain variable $o$ over the ensemble must reasonably be a statistical average of the expectation
values of the same variable over the sub-ensembles \cite{McW1960}. The statistical weight of each addend will naturally be the particles fraction 
(for intensive variables) or the particle number (for extensive variables) of the sub-ensemble. 
For any observable $o$, this supposition implies
\begin{equation}
\begin{array}{l c l}
\textrm{E}\left(o\right)_{\chi}&\equiv&\ddsumd{i=1}{N}\,W_i\, \bra{\beta_i}\hat{O}\ket{\beta_i}   \\
&=& \ddsum{j} \bra{a_j} \left(\ddsumd{i=1}{N} \,W_i\,\ket{\beta_i}\bra{\beta_i}\right)\hat{O}\ket{a_j} \;,
\end{array}
\label{defStatOp}
\end{equation}
where $\hat O$ is related to $o$ as in equation (\ref{eigen}), $W_i$ is the fraction or the number of particles 
of the $i$th sub-ensemble and $\ket{a_j}$ 
are states which form a complete basis in 
the quantum space where $\ket{\beta_i}$ are defined. The relation $\sum_j\ket{a_j}\bra{a_j}=\hat{\mathds{1}}$ has been used in the last step.\\
Now, from equation (\ref{defStatOp}), if $\hat O$ does not depend on the sub-ensembles' states $\ket{\beta_i}$, we clearly see that what depends 
on the characteristics of the ensemble is enclosed in parentheses.
In this case, such quantity can be
assigned to be representative for the ensemble, similarly to the ket vector for a particle state,
and is therefore assigned the name `density operator'.
As a consequence of this assignment, equation (\ref{defStatOp}) becomes nothing but the explicit form of equation (\ref{eq3}).\\
In order to apply equation (\ref{defStatOp}) to the variance and, consequently, to obtain the variance over an ensemble of states
as an expectation value, we should first find an operator which correctly represents
the variance in quantum mechanics. For this purpose, we notice that the correct value for the variance of the spin along the 
$\hat{x}$ direction, over an ensemble of $N$ particles whose states are $\ket{\beta}$, can be obtained as
\begin{equation}
\textrm{Var}(S_x)_{\beta}=N \bra{\beta}\hat{O}_{\hat{\xi}}\ket{\beta}=N \bra{\beta}\hat{O}_{\beta}\ket{\beta} ~,
\label{eq:Oaction}
\end{equation}
where the operator $\hat{O}_{\hat{\xi}}$ reads
\begin{equation}
\begin{array}{l c l}
\hat{O}_{\hat{\xi}}&=&  \left(  \hat{S}_x - E\left( S_x \right)_{\hat{\xi}}    \right)^2 \\[0.3cm]
&=&\ds
\frac{\hbar^2}{4} + \left( E(S_x)_{\hat{\xi}}\right)^2 - 2 \hat{S}_x E(S_x)_{\hat{\xi}}\,.
\end{array}
\label{Otodefine}
\end{equation}
The operator $\hat{\xi}$, as evident from equation (\ref{eq:Oaction}), is defined such that,
by acting on any state $\ket{\beta}$,
it becomes the index which identifies the state. The evaluation of the expectation value contained in the definition of $\hat{O}_{\hat{\xi}}$
is then performed over that state. \\
We could therefore attempt to associate, in quantum mechanics, the operator $\hat{O}_{\hat{\xi}}$
to the variance of the spin along the $\hat{x}$ direction.
However, the definition of the operator $\hat{O}_{\hat{\xi}}$, as given in equation (\ref{Otodefine}),
leads to contradictions. For instance, it follows from equation (\ref{Otodefine}) that the eigenstates of the operator $\hat{O}_{\hat{\xi}}$ would be 
only the two states $\ket{S_x, \pm1}$,
for which the correspondent eigenvalues would be vanishing. Because of this, the operator $\hat{O}_{\hat{\xi}}$ would then be
equivalent to the null operator in quantum mechanics, though the expectation value on $\hat{O}_{\hat{\xi}}$ would not be vanishing over some states, 
such as $\ket{S_z, \pm1}$ or $\ket{S_y, \pm1}$. This is certainly a nonsense! \\
The problem may be identified in the definition of the operator $\hat{O}_{\hat{\xi}}$. The definition of the 
operator $\hat{O}_{\hat{\xi}}$, as given in equation (\ref{Otodefine}), 
is certainly atypical as it does not allow the action of the operator $\hat{O}_{\hat{\xi}}$ on a given 
ket state $\ket{\beta}$ to be disentangled from the statistical evaluation
of the variable $S_x$ over an ensemble of states $\ket{\beta}$.\\
In short, we can say that it seems not possible to assign a well defined quantum operator to the variance and, consequently, 
equation (\ref{defStatOp}) together with the definition of density operator cannot be used to predict the variance for an ensemble of states. 
This is the reason of the wrong prediction
that the statistical formalism has provided for the variance of the spin along the $\hat x$ direction in section \ref{sec:two}.\\
The above discussion strengthens the fact that the description of ensembles of states as given 
by their density operators should be considered incomplete,
as it does not allow the description of some statistical measurable quantities of the ensembles, like the variance.

\medskip

To conclude, we briefly clarify some issues related to unpolarized ensembles.
Since the two ensembles considered in this paper are both unpolarized and have been shown to behave 
differently in experiments, our reasoning raises the 
question as to which ensemble to theoretically consider when unpolarized particles or systems are the object of experiments. 
However, as explained above, the expectation value of physical observables, for which a quantum operator can be safely assigned, 
does not change when the 
unpolarized ensemble is changed, but rather it is equally well described by
any unpolarized ensemble. Since such quantities represent what is normally aimed to be measured in experiments, the problem of choosing the 
right unpolarized ensemble of states is without foundation.
On the contrary, when a full description of the ensemble is required, including observables like the variance, 
then the choice of the unpolarized ensemble to theoretically consider should be constrained by the
information concerning the preparation of the ensemble. 
When the experimental preparation of the ensemble is not under control, the unpolarized ensemble to theoretically consider should be
averaged over the possible representations, or, which is effectively the same, the phase which determines the used representation should be
randomly defined, as pointed out by Tolman in the late thirties \cite{TOLMAN}. \\
In the forthcoming papers, we will further
investigate how the theoretical and experimental results depend on the preparation of the ensemble.

\section{Summary}
\label{sec:sum_conc}
In summary, 
we have showed that ensembles of states which share the same density operator in quantum mechanics, i.e. the same
density matrix, can behave differently in experiments.
In order to prove this statement, 
we started out by defining two ensembles of spin states and by  
showing, without using any quantum statistical mean, that these two ensembles are characterized by a measurable difference,
which is the variance of the spin along a given direction. Then, we moved to analyze the two ensembles within 
the statistical formalism of quantum mechanics.
Since, in this formalism, the two ensembles turn out to be identically described, the variance
of the spin is found to be equal for both of them. The contradiction is then solved out by showing that
the statistical formalism of quantum mechanics cannot be applied at all
to evaluate the variance of the spin of ensembles.
We therefore concluded that the description of ensembles given by their density operators (or by their density matrices)
should be considered incomplete,
as it cannot be applied to predict some measurable statistical quantities and, furthermore, 
as it lacks for measurable information given by the preparation of the ensembles.


\newpage


\section*{References}

\begin{thebibliography}{21}

\bibitem{Neu1927} Von Neumann J 1927 {\it Naturwissenschaften} {\bf 245}
\bibitem{Dir1929} Dirac P A M 1929 {\it Proc. Cambridge Phil. Soc.} {\bf 25} 62
\bibitem{Lan1927} Landau L D 1927 {\it Z. Phys.} {\bf 45} 430
\bibitem{WU} Wu L A, Sarandy M S and Lidar D A 2004 {\it Phys. Rev. Lett.} {\bf 93} 250404
\bibitem{PER} L\"owdin P O 1955 {\it Phys. Rev.} {\bf 97} 1474
\bibitem{WOOTERS} Wootters W K 1998 {\it Phys. Rev. Lett.} {\bf 80} 2245
\bibitem{FRA} Fratini F, Tichy M C, Jahrsetz T, Buchleitner A, Fritzsche S and Surzhykov A 2011 {\it Phys. Rev.} A {\bf 83} 032506
\bibitem{ND}  Dombey O 1969 {\it Rev. Mod. Phys.} {\bf 41} 236
\bibitem{FANO} Fano U 1957 {\it Rev. Mod. Phys.} {\bf 29} 74
\bibitem{McW1960} McWeeny R 1960 {\it Rev. Mod. Phys.} {\bf 32} 335
\bibitem{BLUM} Blum K 1996 {\it Density Matrix Theory and Applications} (New York: Plenum Publishing Corporation)
\bibitem{BAL} Balashov V V, Grum-Grzhimailo A N and Kabachnik N M 2000 {\it Polarization and Correlation
Phenomena in Atomic Collisions} (New York: Kluwer Academic Plenum Publishers)
\bibitem{Bloch:32} Bloch F 1932 {\it Z. Phys.} {\bf 74} 295
\bibitem{Avakyan:07} Avakyan R M, Hayrapetyan A G, Khachatryan B V and Petrosyan R G 2007 {\it Phys. Lett.} A {\bf 372} 77
\bibitem{Glauber:63} Glauber R J 1963 {\it Phys. Rev. Lett.} {\bf 10} 84
\bibitem{Glauber:63b} Glauber R J 1963 {\it Phys. Rev.} {\bf 131} 2766
\bibitem{StatData} Cowan G 1998 {\it Statistical Data Analysis} (Oxford: Clarendon Press)
\bibitem{DIR} Dirac P A M 1982 {\it The Principles of Quantum Mechanics} (Oxford: University Press)
\bibitem{SAK} Sakurai J J 1994 {\it Modern Quantum Mechanics} (Addison-Wesley)
\bibitem{SG} Gerlach W and Stern O 1922 {\it Z. Phys.} {\bf 9} 353
\bibitem{TOLMAN} Tolman R C 1938 {\it The Principles of Statistical Mechanics} (Oxford: Clarendon Press) p 344



\end{thebibliography}
\end{document}